\documentclass[twoside,twocolumn,english,aps, prl, twocolumn, superscriptaddress]{revtex4}
\usepackage[latin9]{inputenc}
\usepackage{graphicx}
\usepackage{amssymb}
\usepackage{esint}

\makeatletter

\providecommand{\LyX}{L\kern-.1667em\lower.25em\hbox{Y}\kern-.125emX\@}

\usepackage{babel}
\makeatother

\begin{document}

\title{Quantum Zeno Effect in the Quantum Non-Demolition Detection of Itinerant
Photons}

\author{Ferdinand Helmer}

\affiliation{{\small Department of Physics, Center for NanoScience, and Arnold-Sommerfeld
Center for Theoretical Physics, Ludwig-Maximilians-Universität, Theresienstrasse
37, D-80333 Munich, Germany}}

\email{Ferdinand.Helmer@physik.lmu.de}

\author{Matteo Mariantoni}

\affiliation{{\small Walther-Meissner-Institut, Bayerische Akademie der Wissenschaften,
Walther-Meissner-Strasse 8, D-85748 Garching, Germany}}

\author{Enrique Solano}

\affiliation{{\small Department of Physics, Center for NanoScience, and Arnold-Sommerfeld
Center for Theoretical Physics, Ludwig-Maximilians-Universität, Theresienstrasse
37, D-80333 Munich, Germany}}

\author{Florian Marquardt}

\affiliation{{\small Department of Physics, Center for NanoScience, and Arnold-Sommerfeld
Center for Theoretical Physics, Ludwig-Maximilians-Universität, Theresienstrasse
37, D-80333 Munich, Germany}}

\begin{abstract}
We analyze the detection of itinerant photons using a quantum non-demolition
(QND) measurement. We show that the backaction due to the continuous
measurement imposes a limit on the detector efficiency in such a scheme.
We illustrate this using a setup where signal photons have to enter
a cavity in order to be detected dispersively. In this approach, the
measurement signal is the phase shift imparted to an intense beam
passing through a second cavity mode. The restrictions on the fidelity
are a consequence of the Quantum Zeno effect, and we discuss both
analytical results and quantum trajectory simulations of the measurement
process. 
\end{abstract}
\maketitle
\newcommand{\ket}[1]{\left|#1\right\rangle }

\newcommand{\bra}[1]{\left\langle #1\right|}

\newcommand{\s}{\hat{\sigma}}

\def\+{\;} 
\def\-{\!}

Quantum non-demolition (QND) measurements are ideal projective measurements
that reproduce their outcome when repeated \citep{1980_Science_Braginsky_QND_Measurement,1992_BraginskyKhalili_QuantumMeasurement}.
Using them, it is possible to measure the state of a system with the
minimal disruption required by quantum mechanics. Recent successful
experimental demonstrations of QND detection for superconducting qubits
and microwave photons \citep{2005_08_Wallraff_PRL_UnitVisibility,2007_Yale_Nature_Photon_Number_Splitting,2007_Gametta_PRA_Optimal_Readout_Quantum_Jump,2008PhRvA..77a2112G}
are both of fundamental interest and crucial for the development of
quantum communication and information processing. When QND detection
is applied continuously to a system that would otherwise undergo some
intrinsic dynamics, quantum jumps are observed, tracing the quantum
evolution in real-time \citep{1993_Cirac_Blatt_PRL_Quantum_Jumps_Ion_Trap,2007_Haroche_Nature_Quantum_Jumps_of_light,1986_Nagourney_Dehmelt_Quantum_Jumps,1987_Putterman_Porrati_PRA_Null_Measurement,1999_PRL_Peil_Gabrielse_QND}.
As a consequence, the dynamics tends to be frozen, a result now known
as the Quantum Zeno effect \citep{1977_J_Math_Phys_Quantum_Zeno,1990_Wineland_PRA_Quantum_Zeno,2007_Blencowe_PRL_Continuous_MEasurement_HO,2007_Rossi_Arxiv_Quantum_Zeno_Cavity_QED_Imperfect_Detectors,damborenea2002mba,2007quant.ph..1007W}.
\begin{figure}
\includegraphics[width=\columnwidth]{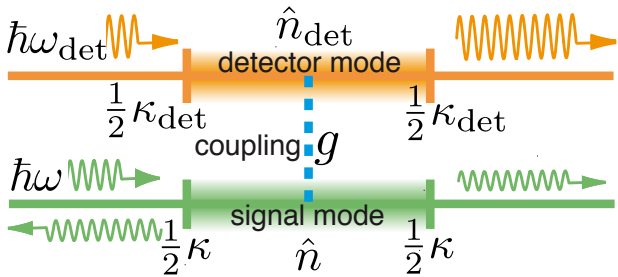}

\caption{(Color online) Schematic sketch of the model. Two cavity modes are
coupled anharmonically. The detector mode is irradiated with a strong
coherent field that suffers a phase shift whenever a photon is present
in the signal mode.}
\label{fig:Schematic_Transport_Situation}
\end{figure}

In the present paper, we show that the interplay of these phenomena
may put interesting constraints on the detection of \emph{itinerant}
quanta. The specific minimal example we will discuss concerns the
continuous dispersive QND detection of single photons passing through
a cavity. The crucial distinction to be recognized is the following:
For localized quanta (e.g. a photon already created inside a cavity
\citep{2007_Haroche_Nature_Quantum_Jumps_of_light,2004_10_Doherty_QND_FockMechanicalOscillator}),
the Quantum Zeno effect could presumably only enhance the detection
by suppressing the decay. However, this no longer holds for the detection
of itinerant quanta, if we require that our detector is always working
and can detect the quantum without knowing the arrival time in advance.
As we will show, in this case the unavoidable back-action of the measurement
device produces a Quantum Zeno effect, suppressing the fidelity of
measurements. 

We investigate a QND scheme utilizing the non-linear, Kerr-type coupling
\citep{imoto1985qnm,yeh1986esn,bachor1988qnm} of two discrete localized
modes of a bosonic field. The presence of a quantum inside the signal
mode gives rise to a frequency shift of the detection mode, which
can be observed dispersively via the phase shift of a beam transmitted
through that mode (see Fig.~\ref{fig:Schematic_Transport_Situation}).
In turn, the signal mode frequency fluctuates due to the detection
beam's shot noise. As a consequence, the incoming signal photon will
be reflected with a probability that rises with coupling strength
and detection beam intensity. 

This incarnation of the Quantum Zeno effect generates a trade-off
that yields the highest detection efficiency at intermediate coupling
strengths. In that way, such dispersive schemes for itinerant quanta
turn out to be similar to weak measurements using general linear detectors
and amplifiers \citep{1992_BraginskyKhalili_QuantumMeasurement}. 

We proceed as follows: (i) We numerically evaluate quantum jump trajectories
for the phase shift signal in a minimal model of a QND photon detector
and analyze the fraction of detected photons, observing the trade-off
described above. (ii) We interpret these findings using an analytical
approximation. (iii) Finally, we briefly comment on possible experimental
realizations.

\emph{Model}. \textendash{} We consider a system of two cavity modes
with a Kerr-type coupling of strength $g$:\begin{eqnarray}
\hat{H} & = & \hbar\omega\left(\hat{n}+\frac{1}{2}\right)+\hbar\omega_{{\rm det}}\left(\hat{n}_{{\rm det}}+\frac{1}{2}\right)+\nonumber \\
 &  & \hbar g\,\hat{n}\:\hat{n}_{{\rm {\rm det}}}+\hat{H}_{{\rm drive}+{\rm {\rm decay}}}\,.\label{eq:Hamiltonian}\end{eqnarray}
These modes might represent two different electromagnetic field modes
inside an optical or microwave cavity, the modes of two adjacent cavities,
or even two anharmonically coupled modes of a nanomechanical resonator.
Photons in the signal mode (frequency $\omega$, number operator $\hat{n}$)
and the detector mode ($\omega_{{\rm det}}$, $\hat{n}_{{\rm det}}$)
decay by leaking out of the cavity. The anharmonic Kerr-type coupling
arises generically when introducing any nonlinear medium, such as
an atom, a qubit or a quantum dot, into a cavity and has been studied
for the purpose of QND measurements in quantum optics \citep{imoto1985qnm,yeh1986esn,bachor1988qnm}.
It induces a phase shift in the strong detection beam ($\left\langle \hat{n}_{{\rm det}}\right\rangle \gg1$)
upon presence of a signal photon. %
\begin{figure}
\includegraphics[width=\columnwidth]{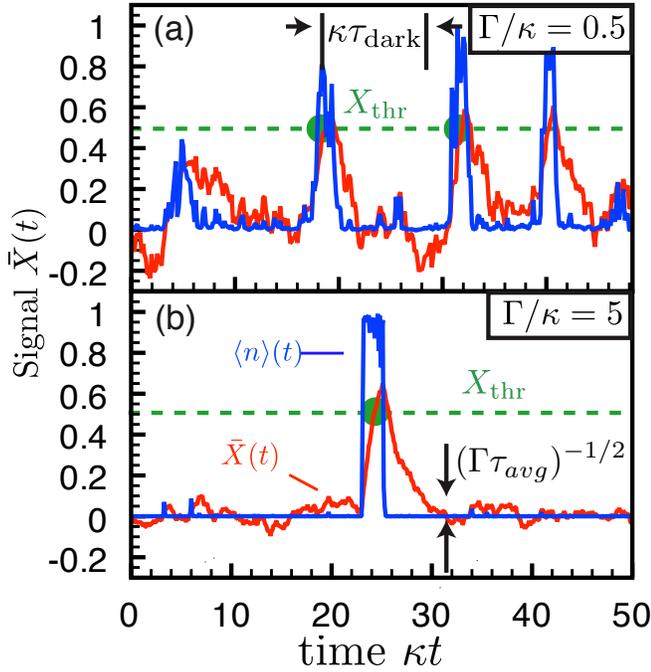}

\caption{(Color online) Quantum jump trajectories illustrating dispersive photon
detection: The observable homodyne signal $\bar{X}(t)$ (red lines)
and the corresponding signal mode occupation $\left\langle \hat{n}\right\rangle (t)$
(blue lines), for two different values of the measurement rate $\Gamma/\kappa$
at a fixed input rate $\dot{N}_{{\rm in}}$. Photon detection events
are indicated as filled circles. The relative noise strength $\propto\left(\Gamma\tau_{{\rm avg}}\right)^{-1/2}$
is suppressed with increasing $\Gamma/\kappa$, but the number of
photons actually detected also decreases, due to the Quantum Zeno
effect (see main text). The size of the noise floor, the detector
threshold $X_{{\rm thr}}$ (green dashed line), and the dark time
are indicated. Here and in the following plots $\kappa\tau_{{\rm avg}}=2$.}
\label{fig:exampletraces}
\end{figure}

We are interested in analyzing individual realizations of the phase
shift signal as a function of time. The phase shift can be observed
by continuously measuring an appropriate field quadrature of the detection
beam (e.g. in a homodyne setup). As the beam passes through the cavity,
the beam becomes entangled weakly with the cavity's state. Thus, the
stochastic measurement outcomes reveal information about that state,
feeding back into the time-evolution of the cavity's density matrix.
This physics is described by a stochastic Lindblad master equation
\citep{1986_PRA_Barchielli_Quntum_Trajectories,1992_BraginskyKhalili_QuantumMeasurement,1993_Carmichael_Open_Quantum_Systems,1995_Walls_Milburn_QuantumOpticsBook,2004_10_Doherty_QND_FockMechanicalOscillator}
for the density matrix $\hat{\rho}$ conditioned on the output signal
(see \citep{2004_10_Doherty_QND_FockMechanicalOscillator}):

\begin{eqnarray}
\dot{\hat{\rho}} & = & -i\sqrt{\frac{\dot{N}_{{\rm in}}\kappa}{2}}\left[\hat{a}+\hat{a}^{\dagger},\hat{\rho}\right]+\kappa\left(\hat{a}\hat{\rho}\hat{a}^{\dagger}-\frac{1}{2}\hat{n}\hat{\rho}-\frac{1}{2}\hat{\rho}\hat{n}\right)\nonumber \\
 &  & -2\Gamma\left[\hat{n},\left[\hat{n},\hat{\rho}\right]\right]-\sqrt{4\Gamma}\left(\hat{n}\hat{\rho}+\hat{\rho}\hat{n}-2\hat{\rho}\left\langle \hat{n}\right\rangle (t)\right)\xi(t).\,\,\,\,\,\label{eq:masterequation}\end{eqnarray}
We analyze a situation with a continuous weak coherent beam of photons
entering at a rate $\dot{N}_{{\rm in}}$ into the signal mode, whose
decay-rate is $\kappa$ (first line of Eq. (\ref{eq:masterequation})).
We have chosen to work in the limit of a large detector mode decay
rate, $\kappa_{{\rm det}}\gg\kappa$, which is favorable for the detection
process and makes it possible to adiabatically eliminate that mode
\citep{2004_10_Doherty_QND_FockMechanicalOscillator}, keeping only
the signal mode $\hat{n}=\hat{a}^{\dagger}\hat{a}$ and drastically
reducing the numerical effort. After adiabatic elimination, the coupling
strength $g$ and the detection beam intensity are combined into the
measurement rate \citep{2004_10_Doherty_QND_FockMechanicalOscillator}
$\Gamma\equiv g^{2}\left\langle \hat{n}_{{\rm det}}\right\rangle /(4\kappa_{{\rm det}})$,
where $1/\Gamma$ is the time-scale needed to resolve different photon
numbers. The last, stochastic term in Eq.~(\ref{eq:masterequation}),
describes the measurement back action. The (suitably normalized) phase
shift signal reads: 

\begin{equation}
X(t)\equiv\langle\hat{n}\rangle(t)+\frac{1}{4}\sqrt{\frac{1}{\Gamma}}\xi(t).\label{eq:xoft}\end{equation}
It contains a systematic term depending on the average number of signal
photons, as well as a stochastic term representing the unavoidable
vacuum noise, where $\left\langle \xi\right\rangle =0$ and $\langle\xi(t)\xi(t')\rangle=\delta(t-t')$.
In deriving Eq.~(\ref{eq:xoft}), we have assumed that the transmitted
and reflected signals are superimposed symmetrically to extract the
maximum information. As in any measurement of field quadratures, temporal
filtering is required to suppress the noise. We average over a timespan
$\tau_{{\rm avg}}$, which should be as large as possible while still
remaining smaller than the expected temporal extent of the phase shift
signal due to a single photon, i.e. $\tau_{{\rm avg}}\ll\kappa^{-1}$.
We denote the averaged signal as $\bar{X}(t)$. 

\emph{Numerical Results}. \textendash{} We numerically solve the master
equation, using it to compute the signal $\bar{X}(t)$ and the occupation
of the signal mode $\left\langle \hat{n}\right\rangle (t)$ as a function
of time. We then implement the minimal model of a threshold detector:
Time-points when the quantum jump trajectory $\bar{X}(t)$ first exceeds
the threshold $X_{{\rm thr}}$ are counted as detection events, and
the detector is then set insensitive for a dark-time $\tau_{{\rm dark}}$,
suitably chosen to avoid multiple detection, i.e. $\dot{N}_{{\rm in}}^{-1}\gg\tau_{{\rm dark}}\gg\kappa^{-1}$.
Our discussion will focus on small values of $\dot{N}_{{\rm in}}$,
making the results independent of $\tau_{{\rm dark}}$, while we will
analyze the dependence on $X_{{\rm thr}}$ in some detail. In figure
\ref{fig:exampletraces} we show two example trajectories. Whereas
the expected number of signal photons is the same for both cases,
the increase in the measurement rate $\Gamma/\kappa$ decreases the
number of photons actually detected, while at the same time enhancing
the signal to noise ratio in the trajectory. This is an indication
of the Quantum Zeno effect in the detection of itinerant quanta, which
we now want to study in more quantitative detail. 

\begin{figure}
\includegraphics[width=\columnwidth]{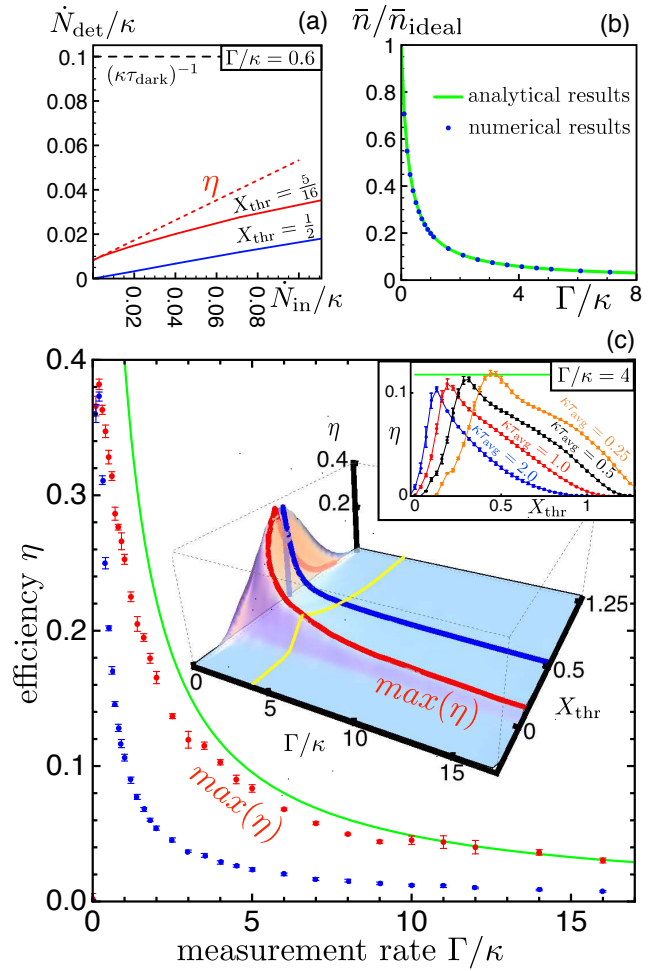}

\caption{(Color online) (a) Detector profile: Rate of detected vs. incoming
photons, at $\Gamma/\kappa=0.6$, for two different thresholds $X_{{\rm thr}}$.
Observe the dark count rate (offset at $\dot{N}_{{\rm in}}=0$), the
detector efficiency $\eta$ defined from the slope at $\dot{N}_{{\rm in}}=0$,
and the saturation for large $\dot{N}_{{\rm in}}\sim\tau_{{\rm dark}}^{-1}$.
(b) Suppression of the signal photon number inside the cavity as a
function of measurement rate $\Gamma/\kappa$. (c) Detector efficiency
$\eta$, obtained from quantum trajectory simulations, as a function
of $\Gamma$ and $X_{{\rm thr}}$ (3D-inset), and comparison with
the analytical results (main plot). The blue data points display $\eta$
for fixed $X_{{\rm thr}}=0.5$ (blue cut in 3D-inset). When maximizing
the efficiency over $X_{{\rm thr}}$ for any given $\Gamma/\kappa$,
the red data points are obtained (labeled {}``$\max(\eta)$''),
coinciding with the analytical asymptote (green thin line) at higher
values of $\Gamma/\kappa$. Small inset: The dependence of the efficiency
$\eta$ on the averaging time $\tau_{{\rm avg}}$ is shown for $\Gamma/\kappa=4$.}

\label{fig:ext_eff_2}
\end{figure}

We plot the rate of photon detection events $\dot{N}_{{\rm det}}$
versus the rate of incoming photons $\dot{N}_{{\rm in}}$ {[}Fig.~\ref{fig:ext_eff_2}(a)].
The detection efficiency $\eta$ is naturally defined as the ratio
of detected vs. incoming photons, obtained at small input rates $\dot{N}_{{\rm in}}$:\begin{equation}
\eta\equiv\left.\frac{d\dot{N}_{{\rm det}}}{d\dot{N}_{{\rm in}}}\right|_{{\rm \dot{N}}_{{\rm in}}=0}.\label{eq:definition_eta}\end{equation}
Fig.~\ref{fig:ext_eff_2}(c) displays the efficiency $\eta$ as a
function of $\Gamma/\kappa$ and $X_{{\rm thr}}$. The statistics
for this figure were obtained from extensive numerical simulations
by generating $O(10^{4})$ trajectories of length $10^{2}/\kappa$
for seven different rates $\dot{N}_{{\rm in}}$ at each value of $\Gamma/\kappa$.
Apparently, the detector efficiency $\eta$ is strongly suppressed
both for $\Gamma/\kappa\ll1$ (low signal to noise ratio) and $\Gamma/\kappa\gg1$.

\emph{Analytical Results}. - To interpret these results, we now calculate
the total transmission probability through the signal mode, whose
frequency fluctuates due to the shot noise in the detection mode,
which is treated as classical noise. We start from the semiclassical
equation of motion for the complex field amplitude $\alpha(t)$ in
the signal mode,\begin{equation}
\dot{\alpha}(t)=\left(-i\,\delta\omega(t)-\frac{\kappa}{2}\right)\alpha(t)+\sqrt{\frac{\kappa}{2}}\alpha_{{\rm {\rm L}}}^{{\rm in}}.\label{eq:EOM_alpha}\end{equation}
Here $\alpha_{{\rm L}}^{{\rm in}}$ is the amplitude of the signal
photon field entering the cavity from the left side, and $\delta\omega(t)\equiv gn_{{\rm det}}(t)$
is the fluctuating frequency shift ($n_{{\rm det}}\gg1$). The correlator
of the noise is given by

\begin{equation}
\left\langle \delta\omega(t)\delta\omega(0)\right\rangle -\left\langle \delta\omega\right\rangle ^{2}=g^{2}\bar{n}_{{\rm det}}e^{-\kappa_{{\rm det}}|t|/2}\,.\label{eq:domcorr}\end{equation}
To obtain an expression for the transmission probability, we write
down the formal solution for $\alpha(t)$,\begin{equation}
\frac{\alpha(t)}{\sqrt{\kappa_{L}}\alpha_{{\rm L}}^{{\rm in}}}=\int_{-\infty}^{t}dt'\exp\left[-i\int_{t'}^{t}\delta\omega(t'')dt''-\frac{\kappa}{2}(t-t')\right].\label{eq:EOM_alpha_solution}\end{equation}
Note that the fluctuations $\delta\omega(t)$ themselves are non-Gaussian.
Still, the integral in the exponent is approximately Gaussian for
time-intervals that fulfill $\kappa_{{\rm det}}|t-t'|\gg1$, due to
the central limit theorem. These times yield the main contribution
under our assumption of a {}``fast detector'', $\kappa_{{\rm det}}\gg\kappa$.
Thus, we can evaluate $\left\langle |\alpha|^{2}\right\rangle $ using
the formula $\langle\exp[-iY]\rangle=\exp[-i\langle Y\rangle-\frac{1}{2}{\rm Var}Y]$
for a Gaussian random variable $Y$, and inserting Eq.~(\ref{eq:domcorr}).
From this, we obtain the average transmitted intensity \begin{equation}
\langle|a_{{\rm {\rm R}}}^{{\rm out}}|^{2}\rangle=\frac{\kappa}{2}\langle|\alpha|^{2}\rangle=\left\langle \mathcal{T}\right\rangle |\alpha_{{\rm L}}^{{\rm in}}|^{2},\label{eq:rel_alpha_in_alpha_out}\end{equation}
and the average transmission probability

\begin{equation}
\langle\mathcal{T}\rangle=\left(1+4\frac{\Gamma}{\kappa}\right)^{-1}.\label{eq:Prob(T)}\end{equation}
Before we can correlate the suppression of the transmission with the
reduction of the detector efficiency $\eta$ in the limit of $\Gamma/\kappa\gg1$,
one more consideration is necessary. In this limit, any photon that
has entered the cavity will almost certainly be detected. Once detected,
the photon loses the coherence with the incoming beam, which is needed
for perfect transmission on resonance in the ideal, coherent case.
As a consequence, it acquires an \emph{equal} probability to leave
the cavity through the left or the right port. This means that, on
average, the number of detected photons is \emph{twice} the number
of transmitted photons. The expected relation is thus $\eta=2\langle\mathcal{T}\rangle$,
which is indeed observed nicely when comparing with the numerical
data (Fig.~\ref{fig:ext_eff_2}(c)). The reduction of detector efficiency
at $\Gamma/\kappa\gg1$ thus has found its explanation in the Quantum
Zeno effect: many photons remain undetected, because they are reflected
due to detector back-action. As low values of $\Gamma/\kappa$ are
also unfavorable, due to a bad signal-to-noise ratio, the maximum
efficiency is found at an intermediate value, namely near $\Gamma/\kappa=1/2$
(see Fig.~\ref{fig:ext_eff_2}).

\emph{Possible realizations}. - Cavity QED setups in superconducting
circuits \citep{2004_09_WallraffEtAl_MicrowaveCavity,2007_Yale_nature_Single_Photon_Source,1977_J_Math_Phys_Quantum_Zeno,2006_PRL_Johansson_96_127006}
have been used to implement ideas of quantum optics on the chip, and
are considered a promising candidate for scalable, fault tolerant
quantum computing \citep{2007_Helmer_cavityGrid_arxiv}. While proposals
for generating nonclassical photon states exist or have been implemented
\citep{2005_Storcz_Arxiv_On-chip_Fock_States,2006_05_Marquardt_PDC,2007_Yale_nature_Single_Photon_Source,2007_Yale_Nature_Photon_Number_Splitting},
the on-chip single-shot detection of itinerant photons is still missing.
Building on recent experiments that demonstrated dispersive qubit
detection \citep{2005_08_Wallraff_PRL_UnitVisibility} and measurements
of photon statistics \citep{2007_Yale_Nature_Photon_Number_Splitting},
one could employ the superconducting qubit to induce a nonlinear coupling
between two modes of the microwave transmission line resonator (or
coupling two cavities %
\footnote{A. Blais, J. Gambetta, C. Cheung, R. Schoelkopf, and S. M. Girvin,
manuscript in preparation (priv. communication)%
}), thus creating a dispersive photon detector of the type discussed
here. These experiments realize a Jaynes-Cummings coupling between
qubit and resonator of up to $2\pi\cdot100{\rm MHz}$, resonators
with frequencies of about $2\pi\cdot5{\rm GHz}$, and a large spread
of resonator decay rates $\kappa$ between $1{\rm MHz}$ and $100{\rm MHz}$.
Given this wide parameter range, it is possible to cover the full
range of $\Gamma/\kappa$ explored here, assuming around $10$ to
$100$ photons in the detection mode. The detector efficiency, although
limited by the Quantum Zeno effect as shown before, could then reach
values of about $40\%$ even without considering more elaborate detector
and signal analysis schemes. Another experiment in which essentially
the same physics could be observed is the detection of single photons
in a microwave cavity by employing the dispersive interaction with
a stream of Rydberg atoms \citep{2007_Haroche_Nature_Quantum_Jumps_of_light}.

\emph{Conclusions}. - In this paper we have analyzed a rather generic
scheme for the detection of itinerant photons in a QND measurement
process, employing quantum trajectory simulations. We have shown how
the Quantum Zeno effect enters the detection efficiency, a result
that will be relevant to many other situations, such as the detection
of electrons tunneling through a quantum dot by current passing through
a nearby quantum point contact \citep{1998_02_Heiblum_WhichPath},
the detection of itinerant phonons entering a micromechanical cantilever
or membrane (e.g. in an optomechanical setup \citep{2008Natur.452...72T}),
and other similar settings in mesoscopic physics, quantum optics,
and atomic physics.

\emph{Acknowledgements}. - We thank S.~M.~Girvin, J.~M.~Gambetta,
A.~A.~Houck, M.~Blencowe, A.~Blais and J.-M.~Raimond for discussions.
Support from the SFB 631, NIM, and the Emmy-Noether program (F.M.)
of the DFG and EuroSQIP (E.S.) are gratefully acknowledged.

\bibliographystyle{apsrev}
\bibliography{SiPh}

\end{document}